\documentclass[nojss]{jss}

\usepackage{framed}

\author{Gillian M Raab\\University of Edinburgh
   \And Beata Nowok\\University of Edinburgh
   \And Chris Dibben\\University of Edinburgh}
\Plainauthor{Gillian Raab, Beata Nowok, Chris Dibben}

\title{Assessing, Visualizing and Improving the Utility of Synthetic Data}
\Plaintitle{Assessing, Visualizing and Improving the Utility of Synthetic Data}
\Shorttitle{Utility of Synthetic Data}

\Abstract{
The \pkg{synthpop} package for \proglang{R} (\url{https://www.synthpop.org.uk/}) provides tools to allow data custodians to create synthetic versions of confidential microdata that can be distributed with fewer restrictions than the original. The synthesis can be customized to ensure that relationships evident in the real data are reproduced in the synthetic data. A number of measures have been proposed to assess this aspect, commonly known as the utility of the synthetic data.  We show that all these measures, including those calculated from tabulations, can be derived from a propensity score model. The measures will be reviewed and compared, and relations between them illustrated. All the measures compared are highly correlated and some are shown to be identical.
The method used to define the propensity score model is more important than the choice of measure.  These measures and methods are incorporated into  utility modules in the \pkg{synthpop} package that include methods to visualize the results and thus provide immediate feedback to allow the person creating the synthetic data to improve its quality. The utility functions were originally designed to be used for synthetic data objects of class \code{synds}, created by the \pkg{synthpop} function \code{syn()} or \code{syn.strata()}, but they can now be used to compare one or more synthesised data sets with the original records, where the records are \proglang{R} data frames or lists of data frames.
}

\Keywords{\pkg{synthpop}, synthetic, utility, tuning, \proglang{R}}
\Plainkeywords{synthpop, synthetic, utility, R}

\Address{
  Gillian M Raab\\
  School of GeoSciences\\
  University of Edinburgh\\
  Drummond Street\\
  Edinburgh EH8 9XP, Scotland, United Kingdom\\
  E-mail: \email{gillian.raab@ed.ac.uk}\\
  URL: \url{https://www.synthpop.org.uk/}
}

\begin{document}

\section[Introduction]{Overview}\label{sec:intro}

The utility of synthetic data will ultimately be measured by how results from analyses of synthetic data and the conclusions following from them will differ from those derived from the real data. It is not advisable to tune synthesis methods to make the results of a specific analysis agree with those from the original. Details of the final analyses are seldom known and, even if they were, creating the synthesis to give agreement for an analysis model will give answers that will agree, but the residuals from the model fitted to the synthetic data will not give any evidence of model inadequacy that might have been found with the original. Thus more flexible models that will reflect all the relationships in the data are to be preferred \citep[see][Section 5 for a discussion of this]{Raab_jpc}. There is a need for measures that compare wider aspects of the differences between the synthetic and original data to give feedback on the utility of the synthesis. Such measures are termed ``broad'', ``global'' or ``general'' utility measures, as opposed to ``narrow'' or ``specific'' measures that focus on the results of particular analyses. We review the wide range  of general utility measures that have been proposed for synthetic data and the methods used to calculate them. All the measures discussed can be calculated in the \proglang{R} package \pkg{synthpop}. We present examples and recommendations of how they can be used in practice.

There are two main reasons we might wish to evaluate the utility of synthetic data:
\begin{enumerate}
	\item {To compare different synthesis methods for the same data set.}
	\item {To diagnose where the original and synthetic data distributions differ and thus tune the synthesis methods to improve the utility of the synthetic data.}
\end{enumerate}

For both of these reasons we recommend the propensity mean squared error ($pMSE$) as a  utility measure. For the first we advise fitting the propensity score model by a classification and regression tree (CART) model. The default printed output from all the utility functions therefore presents only the $pMSE$ measure and its standardized ratio ($S_pMSE$). All the other utility measures discussed here and listed in Table~\ref{tab:detmeas} in Appendix~\ref{app:detmeas} are also available as outputs from the utility functions.
  
The code below compares two different methods of synthesizing ten variables selected from \code{SD2011}, the survey data that are part of \pkg{synthpop}. The selection is stored in a data frame \code{ods}. The first synthesis uses parametric models\footnote{Setting \code{method = "parametric"} selects models appropriate to the type of variable. The parameter \code{cont.na} sets -8 as a second type of missing value. The method used for numeric variables uses a transformation to the expected Normal ranks so as to preserve the univariate distributions for skewed variables.}, while the second synthesizes from CART models. The parameter \code{cont.na} for the variable \code{income} defines another type of missing value (coded as \code{-8}) to be kept as a separate group. Both the utility evaluations use a pr\textbf{}opensity score model fitted by CART and results show that the utility score is three times higher (= worse) for the parametric synthesis compared to the CART synthesis.

\begin{CodeChunk}
	\begin{CodeInput}
R> library(synthpop)
   ods <- SD2011[, c("sex", "income", "age", "edu" , "socprof", "trust",
   "height", "weight", "smoke", "region")]
R> syn_para <- syn(ods, method = "parametric", cont.na = list(income = -8),
   seed = 34567, print.flag = FALSE)
R> syn_cart <- syn(ods, method = "cart", cont.na = list(income = -8),
   seed = 34567, print.flag = FALSE)
R> u_ods_para <- utility.gen(syn_para, ods, method = "cart",
   resamp.method = "none", print.flag = FALSE)
R> u_ods_cart <- utility.gen(syn_cart, ods, method = "cart",
   resamp.method = "none", print.flag = FALSE)
   cat("\nParametric pMSE ", u_ods_para$pMSE,
   "\nCART pMSE ", u_ods_cart$pMSE,
   "\nUtility ratio parametric to CART: ", u_ods_para$pMSE/u_ods_cart$pMSE)
 
\end{CodeInput}
\begin{CodeOutput}
Parametric pMSE 0.03826426 
CART pMSE 0.008475425 
Utility ratio parametric to CART: 4.51473  
\end{CodeOutput}
\end{CodeChunk}

\begin{figure}[h]
	\centering
	\includegraphics[width=0.9\textwidth]{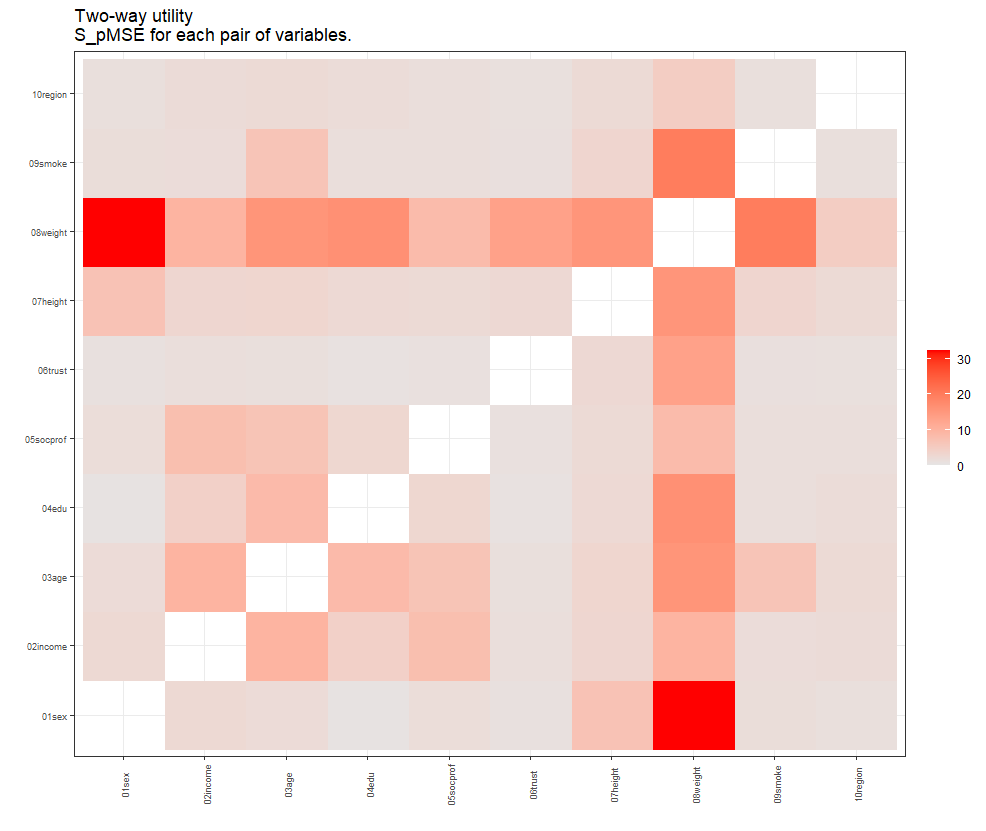}
	\caption{\label{fig:syn_ods1} Plot produced from synthesis of \code{ods} data by parametric methods.}
\end{figure}

To diagnose problem, we recommend visualizing the how well the synthetic data preserves relationships between subsets of the variables, e.g., all one-way, two-way or three-way combinations. The code below visualises utility for all two-way tables from the parametric synthesis, with five groups created from the non-missing values of continuous variables, and $S\_pMSE$ plotted  (Figure~\ref{fig:syn_ods1}). It is clear that the variable with the most problems is \code{weight}. Section~\ref{sec:tuning} will illustrate further syntheses of these data.

\begin{CodeChunk}
	\begin{CodeInput}
R> utility.tables(syn_para,ods, tables = "twoway", nworst = 4)
	\end{CodeInput}
	\begin{CodeOutput}
Two-way utility  utility measure (S_pMSE) plotted for 45 pairs of variables.
Variable combinations with worst 4 utility scores (S_pMSE):
01sex:08weight  08weight:09smoke    04edu:08weight 07height:08weight 
30.95             18.46             15.73             14.63 

Median and maximum of selected utility measures for all tables compared
Median
pMSE S_pMSE     df 
0.00   1.95  28.00 
Maxima
pMSE S_pMSE     df 
0.01  30.95 153.00 
	\end{CodeOutput}
\end{CodeChunk}

These recommendations are based on our practical experiences and on empirical evaluations that are detailed in the rest of this paper. In Section~\ref{sec:meas} we present details of all the utility measures, their performance in evaluating syntheses and the relationships between the measures. We show that two sets of seemingly unrelated utility measures (one pair and one set of three) are identical. Section~\ref{sec:models} evaluates models that can be used to set up the comparison of the real and synthetic data from which the utility measures are derived. Section~\ref{sec:tuning} provides examples of using the utility functions to diagnose problems and tune the synthesis methods to improve utility. The examples also allow us to introduce some new functionality that has been added to the \pkg{synthpop} package since its initial description in \cite{synthpop_1.0-0_CRAN, synthpop_JSS}. The final Section~\ref{sec:conc} summarizes the paper and makes suggestions as to possible future enhancements to these methods.

The functions described in this paper were designed to be used by synthetic data objects of class \code{synds} created by the function \code{syn()} or \code{syn.strata()}. Since Version~1.7-0 of \pkg{synthpop} they also work for a synthetic data set (or a list of several data sets) created by other methods.

\section[Measures]{Choice of utility measures}\label{sec:meas}
One approach to general utility measures involves combining the original and synthetic records and measuring how well the data values can predict the source of the records as real or synthetic \citep{Karr_Oganian_Reiter_Woo_2006, wooProp}. This method uses the propensity score, $\hat{p}$, the predicted probability that a record comes from the synthetic data. If the synthesis has been carried out from a model that is compatible with the original data distribution, then the expected mean of $\hat{p}$ will be $c = n_2/N$, where there are $n_1$ records from the original data and $n_2$ from the synthetic data and $N = n_1 + n_2$. We refer to the  distributions of utility measures in this case as their Null distributions. The most commonly suggested utility measure is known as the propensity score mean square error ($pMSE$). The Null distribution of the $pMSE$, for prediction models with a fixed number of parameters has been derived by \cite{snoke_et_al_2018} and its expectation is $df c(1-c)^2/N$, where $df$ is the number of degrees of freedom constrained by fitting the propensity score model. Other utility measures can also be derived from the propensity score, e.g. the percentage above 50\% of records correctly predicted (PO50) and the Kolmogorov-Smirnov statistic ($SPECKS$) which is the maximum distance between the cumulative distributions functions (CDFs) of the propensity score for the synthetic and original distributions \citep{Bowen_Liu_Su}. Further measures that compare $\hat{p}$ values between the original and synthetic data could be considered. One such is the Wilcoxon signed-rank statistic ($U$). 

An alternative approach to utility measures is to group the original and synthetic data, usually by constructing tables based on their values, and to compute measures of difference between the tables. \cite{Voas_Williamson_2001} investigated measures based on the family of goodness-of-fit measures discussed by \cite{Read_Cressie_1988}. They note that the usual Pearson $\chi^2$ statistic needs to be adjusted because synthetic data may be generated in cells where the count from the original data is zero. They propose replacing the expected value in the denominator of the formula with the average of the original and synthetic counts. This statistic and its generalization when $n_1 \neq n_2$ are designated as $VW$. Other goodness-of-fit measures that can be calculated from tables include the  the Freeman-Tukey statistic ($FT$)\footnote{This measure is proportional to the discrete Hellinger distance between two distributions.}, the Jensen-Shannon divergence ($JSD$) and the likelihood ratio $\chi^2$ statistic ($G$). The likelihood ratio has no contributions from cells where the original counts are zero. 
It would be desirable for these cells to contribute to utility measures since they may be a substantial 
proportion of all cells, especially for sparse tables. Another possible measure derived from tables is the mean of the absolute differences between the distributions of original and synthetic counts, designated as $MabsDD$\footnote{Suggested by Christine Task, as used to evaluate the NIST challenges, see \href{https://www.nist.gov/ctl/pscr/open-innovation-prize-challenges/current-and-upcoming-prize-challenges/2020-differential}{here}.}. A related quantity is $WMabsDD$, where the absolute differences are weighted in proportion to the inverse of the standard deviation of their Null expectations, so that this measure has a known Null expectation. Yet another measure, the Bhattacharyya  distance ($dBhatt$), a measure of histogram overlap \citep{dbhatta}, has been used for comparing tables of original and synthetic data by \cite{significance}.

Corresponding to the two approaches, \pkg{synthpop} provides two functions to calculate utility measures, \code{utility.gen()} and \code{utility.tab()}. 
Comparing tables of original and synthetic data can be framed as a prediction model, where the propensity score for records in each cell is the ratio of the synthetic counts to the sum of the original and synthetic counts. For synthetic data where all variables are categorical, a comparison of $n$-way tables is equivalent to fitting a propensity score model by logistic regression including all interactions up to order $n$. Thus, any measure defined from the propensity score can also be computed for tables, but some tabular utility measures do not correspond to measures from the propensity score approach.

In Table~\ref{tab:detmeas} in Appendix~\ref{app:detmeas} we provide details of all the measures mentioned above; the number of possible measures in the table is not as many as first appears; the $pMSE$ is just a multiple of the tabular utility measure $VW$. $dBhatt$ is a multiple of $\sqrt{FT}$; the three measures $SPECKS$, $PO50$ and $MabsDD$ are linearly related for most examples. The code below 
shows this empirically for the 120 different syntheses from all three-way tables generated by the parametric synthesis, and the proofs of linearity are in Appendix~\ref{app:plot}. 
\begin{CodeChunk}
	\begin{CodeInput}
R> u_3way <- utility.tables(syn_para, ods, 
   utility.stats = "all", tables = "threeway")
R> cors <- cor(u_3way$tab.res)
R> cat("Correlations VW with pMSE =",cors["VW","pMSE"],
   ", SPECKS with MabsDD=",cors["SPECKS","MabsDD"],
   " and SPECKS with PO50=",cors["SPECKS","MabsDD"],".",sep="")     
R> toplot<- u_3way$tab.res[,c(1:4,6,5,7)]
   dimnames(toplot)[[2]][c(1,4)] <- c("VW\npMSE","SPECKS\nMabsDD\nPO50")
  pairs(toplot)   
	\end{CodeInput}
	\begin{CodeOutput}
Correlations VW with pMSE =1, SPECKS with MabsDD=1 and SPECKS with PO50=1	
    \end{CodeOutput}
\end{CodeChunk}

\begin{figure}[h]
	\centering
	\includegraphics[width=0.98\textwidth]{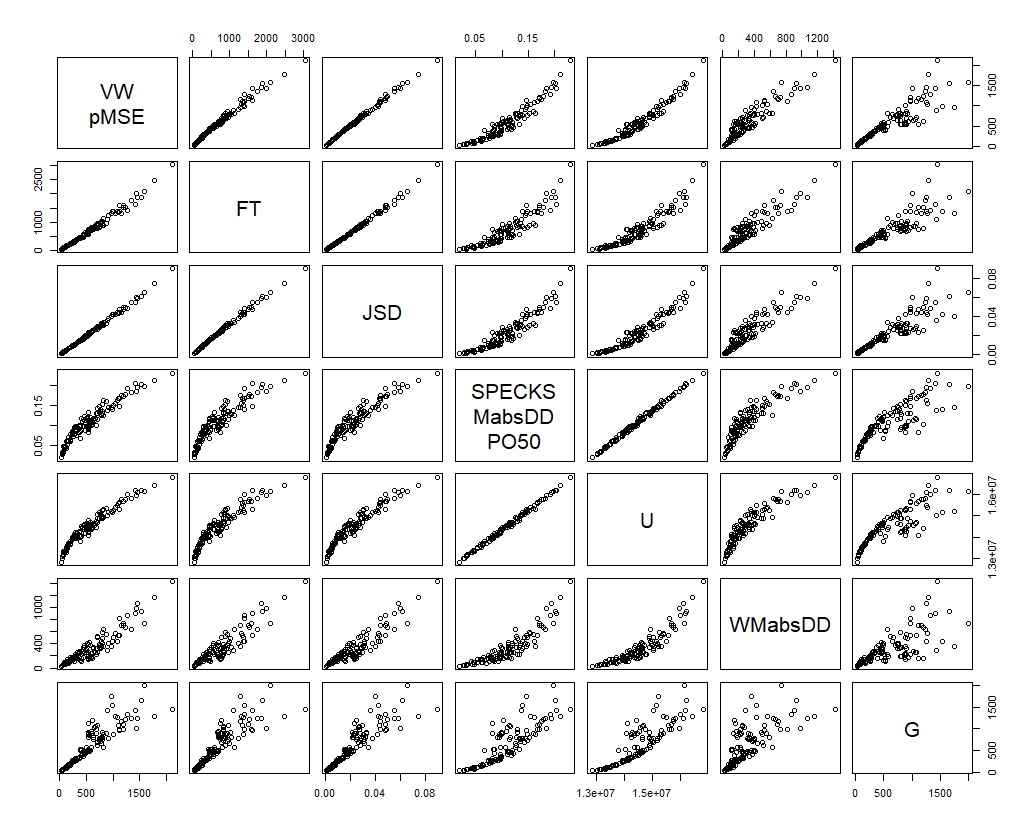}
	\caption{\label{fig:pairs} Pairs plot of utility measures for all 120 three-way tables from the synthesis of \code{ods} data by parametric synthesis.}
\end{figure}

The pairs plot from this code is shown as  Figure~\ref{fig:pairs}. All of  correlations exceeding 0.9, with the exception of those with $G$.  Correlations between $VW:pMSE$, $FT$ and $JSD$,exceed 0.99, as does that for $SPECKS:MabsDD:PO50$ with $U$.

\begin{table}[h]
	\centering
	\begin{tabular}{rrrrrrrrrrrrr}
		\hline
		       & Mean & Mean & Mean & Mean & Mean & Mean  & Mean & Median & Median & Median & Median\\
 & &  &  &  &  & $WMa-$ & &\multicolumn{2}{c}{correct}&\multicolumn{2}{c}{bootstrap}\\
	n  & $VW$ & $FT$ & $JSD$ & $PO50$ & $U$ & $bsDD$ & $G$ & $df$* & $df$* & $dfG$\# & $dfG$\#\\

		\hline
2 & 39.2 & 39.4 & 39.3 & 16.9 & 16.1 & 16.2 & 39.9 & 13 & 13 & 13 & 13 \\ 
3 & 121.5 & 129.1 & 126.8 & 39.3 & 42.8 & 38.1 & 148.6 & 69 & 69 & 69 & 69 \\ 
4 & 143.9 & 145.1 & 149.0 & 62.8 & 65.3 & 58.4 & 169.5 & 274 & 265 & 292 & 267 \\ 
5 & 254.5 & 280.4 & 273.7 & 121.7 & 111.8 & 164.4 & 61.5 & 1011 & 833 & 2178 & 771 \\ 
6 & 240.5 & 266.9 & 258.5 & 127.5 & 114.5 & 185.4 & 26.3 & 1576 & 1229 & 3556 & 792 \\
		\hline
		\multicolumn{12}{l}{* Effective degrees of freedom is one less than the number of cells in the cross-tabulation }\\
		\multicolumn{12}{l}{~~~of all variables that contain any original or synthetic counts. This depends on the synthetic data.}\\
		\multicolumn{12}{l}{\# For computing $G$ only cells with both original and syntheic counts contribute.}\\	
		\hline	
	\end{tabular}
	\caption{\label{tab:power} Empirical power of different utility measures comparing ``incorrect'' with ``correct'' categorical syntheses calculated by \code{utility.tab()} with the code given in Appendix~\ref{app:evalmeas}.} 
\end{table}

Appendix~\ref{app:evalmeas} gives details of a simulation for up to 6 categorical variables of the power of each measure to distinguish between a correct synthesis and one where all the variables are unrelated. The correct synthesis uses a saturated log-linear model, while the incorrect one uses independent bootstrap samples for each variable. Table~\ref{tab:power} summarises the evaluation of the power of all the statistics that can be calculated by \code{utility.tab()}. As expected $pMSE$, $FT$ and $JSD$ all have similar power. The likelihood ratio statistic, $G$, has similar power for tables with large expected counts but loses power for sparse tables.  The other three measures have lower power for these examples. The code to generate the results in Table 1 is given in Appendix~\ref{app:evalmeas}.

One desirable feature of utility measures are that they should be available from propensity score methods not based on tables and another is a known Null expectation. Only $pMSE$ has both of these properties. Another advantage of $pMSE$ and $WMabsDD$ is that they can be standardised from a single synthetic data set.  A standardized measure of $SPECKS$ requires a replication method using multiple synthetic data sets (see Appendix~\ref{app:detmeas}). The measure $U$ has the poorest power for this example, while the power of $G$ deteriorates for large sparse tables.

\section{Models for the propensity score: practical considerations.}\label{sec:models}

As well as choosing a utility measure, the synthesizer must decide on which model is to be used to calculate the propensity score. The two possible classes of models are logistic regression and adaptive classification models such as CART. Within each class, a variety of models can be specified by defining predictors for logistic models and by altering the methods and settings of classification models.  The three models now available in the utility modules of \pkg{synthpop} are given in Table~\ref{tab:models}. All can be computed from \code{utility.gen()} but only one also from \code{utility.tab()}. The choices between these models are largely based on practical considerations, as we discuss below.

\begin{table}[h]
	\centering
	\begin{tabular}{llcc}
		\hline
		Model          & Description & \code {utility.gen()} & \code{utility.tab()} \\
		\hline
		(a) Saturated logistic & Logistic regression with  & x & x \\
		                       & all interactions up to the  \\
		                       & number of variables in data \\
		(b) Logistic to order $n$ & Logistic regression with all & x & \\
		                          & interactions up to order $n$\\

	    (c) CART models & Classification and regression & x &\\
		                & trees & \\
		\hline
	\end{tabular}
	\caption{\label{tab:models}Propensity score models implemented in \pkg{synthpop}.} 
\end{table}

Models of type (a) calculated from \code{utility.tab()} are limited by the memory required to hold large tables and by the fact that large tables can become sparse so that their statistical properties may be uncertain. The six variables contributing to the evaluation of the Null models used to calculate results for Table~\ref{tab:power} defined a table with 14,000 cells, though only 3,500\footnote{Median from 10,000 syntheses} of the cells contain any counts from either the original or synthetic data. A table of all 10 variables in the data set \code{ods} would contain over 14 million cells although only 0.04\% of them would contain any counts. Memory problems would prevent this method from being used for 7 or more variables from this data set, and the sparsity of the tables would advise against using tables of more than 5 variables. To try to fit model (a) via logistic regression does not help either because it is constrained by its large number of parameters. For the first 5 variables from the \code{ods} data set, including all possible interactions, requires a model with 3,500 parameters that failed to converge in several hours of computing time. Thus method (a) can only be used for a few variables at a time.

Logistic models for data sets with many variables are limited by the number of parameters required to fit the propensity score model. Using method (b) with the default setting of all second-order interactions for the  10 variables in \code{ods} gives a model with 753 parameters. This model fitted in under two minutes\footnote{On a Windows laptop with spec to add}. A model with three-level interactions of all variables would have defined a model with over 7,000 parameters. Models with second-order interactions may have problems especially if they contain factors with many levels. The choice of model to fit the propensity score for a data set with many variables is between logistic regression (b), with restrictions on interactions, and a CART model (c). A CART model (c) requires the use of resampling methods if a standardized measure is required, but only a single synthetic data set is required and results seem satisfactory (see Appendix~\ref{app:evalNULL}). We have found that CART models can diagnose differences more easily with fewer computational problems than logistic models.  For a single summary measure to compare syntheses we recommend a CART propensity score model, with $pMSE$ standardised by a permutation method (see the example in Section~\ref{sec:intro}).

\begin{figure}[h]
	\centering
	\includegraphics[width=0.9\textwidth]{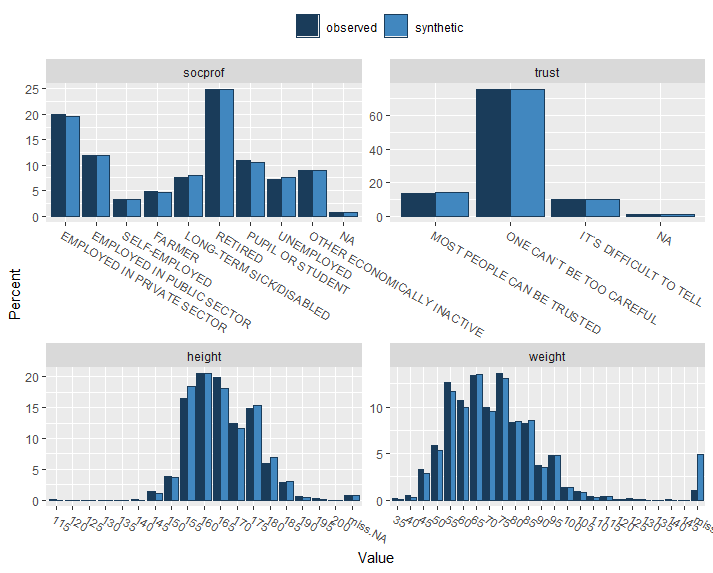}
	\caption{\label{fig:Fig3} Extract from the graphical output of  \code{compare} used to evaluate by parametric synthesisof \code{ods}.}
\end{figure} 
  
The choice of model for diagnosing and fixing problems with a synthesis is different. For this we need a method that will pinpoint the parts of the distribution of synthetic data that differ from that of the original.  It is possible to examine the trees that have been used to calculate the propensity score from CART models or the coefficients of the propensity score model, but in practice the effect of individual variables is difficult to identify.
 
 An approach that is much more practical and useful is to examine the agreement between the synthetic and original data for low-order margins: starting with one-way marginals, then two-way and perhaps three-way. This approach is illustrated in the next section. A first step is to calculate utilities from all one-way marginals and compare plots. Since Version~1.7-0 this function also produces a table of utility measures for each one-way table. The code below gives utility statistics for this example and {\label{fig:Fig3} is an extract from the set of plots produced.

 \begin{CodeChunk}
	\begin{CodeInput}
R> compare(syn_para, ods, utility.for.plot = "S_pMSE" )
	\end{CodeInput}
	\begin{CodeOutput}
Selected utility measures:
         S_pMSE    df
sex      0.5489709  1
income   1.0022429  6
age      1.0549926  4
edu      0.3552980  4
socprof  0.4053007  9
trust    0.8958318  3
height   2.9677025  5
weight  52.6684121  5
smoke    0.5396529  2
region   1.3914925 15
	\end{CodeOutput}
\end{CodeChunk}

The utility measures pinpoint \code{weight} as the problem variable, and the plot identifies the excess of missing values  created in the synthetic data as the problem. 

\section{Using utility measures to tune the synthesis methods.}\label{sec:tuning}

The \pkg{synthpop} package allows the syntheses to be tuned in various ways to adapt to the needs of particular data sets. These include:
	\begin{itemize}
	\item{Changing the order in which the conditional distributions are formed.}
	\item{Stratifying the synthesis by important variables.}
	\item{Changing the methods for individual variables.} 
	\item{Modifying the predictor matrix to exclude certain variables as predictors of others.}

\end{itemize}
 
 The first two are the ones we have found most useful. We have found the need for the third and fourth only in special circumstances, one of which we will describe here. Some survey or administrative data contain very detailed fields that can be grouped into wider classes. Examples are the classification of occupations or diagnostic codes. The detailed variables are nested within the wider one.  The detailed variables have too many classes to be used as predictors. To overcome this they need to be synthesised after the wider class and are given the method \code{"nested"}. This creates synthetic data for the detailed variable by taking bootstrap samples within the groups.  The prediction matrix needs to be modified to remove the detailed variable as a predictor of other variables. Details of this and other possible strategies to improve syntheses are discussed in \cite{Raab_Nowok_Dibben_2017}.
 
 Use of the first two methods is illustrated in the for the synthesis of \code{ods}, first by parametric methods. In all the examples we have specified the default utility value, the standardised  ratio,$S\_pMSE$ calculated from its expectation for logit models and by a permutation method for CART models. The target value for this utility model is 1.0, but we do not believe that real world data is ever generated exactly from a model. Thus we do not calculate any significance tests. We have found that a useful rule for practical use is to aim for utility ratios below 10. 
 
 The first step of evaluating the utility of any synthesis is to compare the univariate distributions for each variable, illustrated for our example in Section~\ref{sec:intro} and \label{fig:Fig3}.
 This type of problem with small categories can happen with variables that are synthesized towards the end of the list of conditional distributions. This is easily fixed by moving the variable \code{weight} up towards the start of the visit sequence.
\begin{CodeChunk}
	\begin{CodeInput}
R> syn_para2 <- syn(ods, method = "parametric", cont.na = list(income = -8),       
+   visit.sequence = c(1, 3, 7:9, 2, 4:6, 10), seed = 34567,
+   print.flag = FALSE)
R> compare(syn_para2, ods, c("S_pMSE","df"),  plot = FALSE )
 	\end{CodeInput}
 	\begin{CodeOutput}
Selected utility measures:
        S_pMSE     df
sex     0.54897088  1
income  0.42940156  6
age     0.87879201  4
edu     3.73195942  4
socprof 2.28987779  9
trust   2.01417616  3
height  3.54288687  5
weight  0.33657160  5
smoke   0.07600835  2
region  1.43653039 15	
	\end{CodeOutput}
\end{CodeChunk}	
 
With this new order it is now time to investigate the two-way relationships between variables for the reordered synthesis. Figure~\ref{fig:syn_ods_4} (a) to (d) shows the default two-way plots from \code{utility.tables()} from four different syntheses. Note that these plots are all scaled to approximately the same legend as was generated by the range of utilities in the first synthesis: Figure~\ref{fig:syn_ods1} reproduced as Figure~\ref{fig:syn_ods_4} (a). Figure~\ref{fig:syn_ods_4} (b) shows the two-way plots from the reordered synthesis, clearly much better, although with some high values, notably those for interactions with \code{age} where there are some utility values above 10. Stratifying the synthesis by dividing into two strata, \code{age} above and below 55, brings the maximum utility ratio down to below 7 (Figure~\ref{fig:syn_ods_4} (c)), but note that had we used CART synthesis with the original ordering  (Figure~\ref{fig:syn_ods_4} (d)) the maximum utility ratio would have been below 3.
 
In this example our preferred CART models did not require any improvement. Large complex data sets, even synthesized by CART, often require strategies mentioned above to improve their utility. Stratifying the synthesis by variables known to be of interest in any planned analyses is a good strategy to ensure relationships will be maintained in the synthetic data. 
 
 \begin{figure}[h]
 	\centering
 	\includegraphics{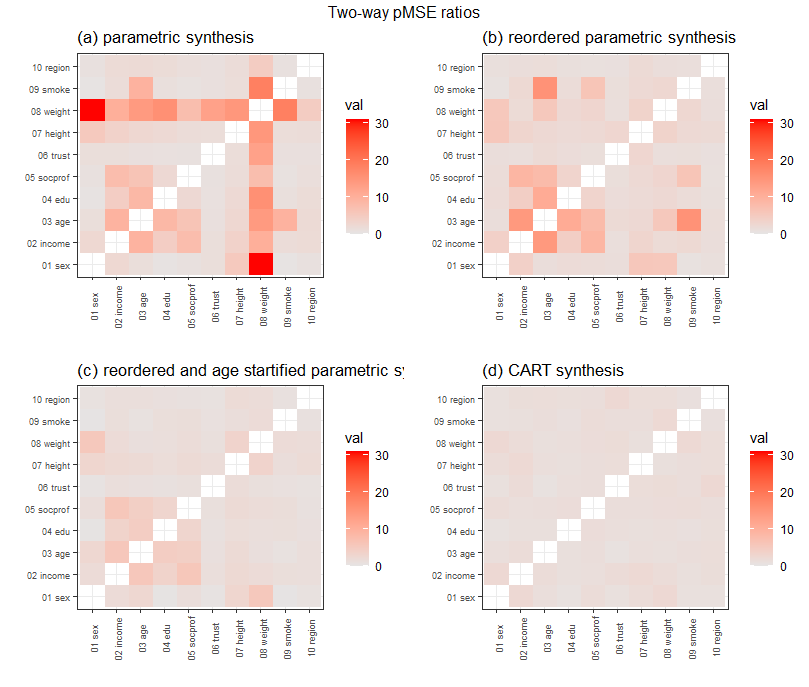}
 	\caption{\label{fig:syn_ods_4} Plots produced from two-way utility measures for synthesis of \code{ods} by parametric synthesis.}
 \end{figure}
 
\begin{CodeChunk}
	\begin{CodeInput}
R> syn_para3 <- syn.strata(ods, method = "parametric", 
+   strata = ods$age > 55 & !is.na(ods$age), cont.na = list(income = -8), 
+   visit.sequence = c(1, 3, 7:9, 2, 4:6, 10), seed = 34567, 
+   print.flag = FALSE) 

R> u.para <- utility.tables(syn_para, ods, 
+   plottitle = "(a) parametric synthesis")
R> u.para2 <- utility.tables(syn_para2, ods, 
+   plottitle = "(b) reordered parametric synthesis")
R> u.para3 <- utility.tables(syn_para3, ods, 
+   plottitle = "(c) reordered and age startified parametric synthesis")
R> u.cart <- utility.tables(syn_cart, ods, 
+   plottitle = "(d) CART synthesis")

R> list.plots <- vector("list", length = 4)
R> list.plots[[1]] <- print(u.para, max.scale = 31)$p
R> list.plots[[3]] <- print(u.para2, max.scale = 31)$p
R> list.plots[[2]] <- print(u.para3, max.scale = 31)$p
R> list.plots[[4]] <- print(u.cart, max.scale = 31)$p
R> gridExtra::marrangeGrob(list.plots, nrow = 2, ncol = 2, 
+   top = "Twoway pMSE ratios")
\end{CodeInput}
	\end{CodeChunk}
The function \code{utility.tables()} can also calculate utility measures for all three-way tables. Plots like Figure~\ref{fig:syn_ods1} are produced for three way tables holding one of the variables fixed. This third variable can be specified by the user. If this is not done then the program selects the variable with the highest utility score over all tables it contributes to.

\section{Conclusion}\label{sec:conc}
This paper started life as a simple ``how-to-do-it'' explanation of the routines we have written to measure data utility. Documenting them all in detail has led to some unexpected insights into the utility measures. What is more, it provides a firm foundation for our rules as to how to proceed to assess the utility of synthetic data and improve its quality. Briefly:
\begin{enumerate}
	\item{To compare the overall utility of two methods of synthesizing the same original data, you should fit a propensity score model with an adaptive model such as CART and compare the $pMSE$ measures, calculated by \code{utility.gen()}, for the two methods.}
	\item{To judge and improve the utility of a synthesis method:}
		\begin{itemize}
	\item {Start by visualizing all the one-way tables with \code{compare()}.} 
	\item{Next visualize all two-way ratios with \code{utility.tables()}}. 
	\item{If all the standardized $pMSE$ ratios are below 10, or better still below 3, it is probably not necessary to do anything more as the utility seems acceptable.}
	\item{At each of the steps above you should try to improve the utility by tuning the synthesis with stratification and/or by changing the default parameters of \code{syn()}.}
	\end{itemize}
		\end{enumerate}
These recommendations  have been exemplified on just one example, but we have found similar results from other data sets. We hope that other \pkg{synthpop} users can try out these functions on their own data and provide feedback on ways we might improve the utility functions and the ways they can be used.

Other measures of differences between the original and synthetic data could also be considered. One such is the discretised earth mover's distance (EMD) or Wasserstein distance, used for synthetic data by \cite{emd}. It measures the cost of transporting the probability mass from one  distribution to make it match the second. It requires a cost function for each pair of cells in the table. If the costs for every pair of cells were the same, then the EMD would just be the same as the $PO50$. A measure that gave different costs would clearly be preferable, especially for ordered categories, but would involve a detailed specification. Suggestions on how this might be achieved would be welcome. Further metrics or methods may also be possible and we would welcome suggestions for these.

Another important aspect of utility is feedback from those to whom the synthetic data are supplied. One example of this was a synthesis we carried out of dates when children were excluded from school. By definition these dates need to be weekdays (although this was not true for a few original records). The synthetic data spread the dates over weekends too. To overcome this the data would need to be pre-processed to define the variables differently. This is a different aspect of utility and examples like this are common and as important as the more formal utility measures discussed here. 

\section*{Acknowledgments}
Thanks to users of \pkg{synthpop} who have provided feedback on the methods used. 

\bibliography{utility}

\begin{thebibliography}{13}
\newcommand{\enquote}[1]{``#1''}
\providecommand{\natexlab}[1]{#1}
\providecommand{\url}[1]{\texttt{#1}}
\providecommand{\urlprefix}{URL }
\expandafter\ifx\csname urlstyle\endcsname\relax
  \providecommand{\doi}[1]{doi:\discretionary{}{}{}#1}\else
  \providecommand{\doi}{doi:\discretionary{}{}{}\begingroup
  \urlstyle{rm}\Url}\fi
\providecommand{\eprint}[2][]{\url{#2}}

\bibitem[{Bhattacharyya(1943)}]{dbhatta}
Bhattacharyya A (1943).
\newblock \enquote{On a Measure of Divergence Between Two Statistical
  Populations Defined by Their Probability Distributions.}
\newblock \emph{Calcutta Mathematical Society}, \textbf{35}, 99--109.

\bibitem[{Bowen \emph{et~al.}(2021)Bowen, Lui, and Su}]{Bowen_Liu_Su}
Bowen CM, Lui F, Su B (2021).
\newblock \enquote{Differentially Private Data Release via Statistical Election
  to Partition Sequentially.}
\newblock \emph{METRON}, \textbf{79}(1), 1--31.
\newblock \urlprefix\url{https://doi.org/10.1007/s40300-021-00201-0}.

\bibitem[{Grosso \emph{et~al.}(2020)Grosso, Pichler, and Piantanida}]{emd}
Grosso GD, Pichler G, Piantanida P (2020).
\newblock \enquote{Privacy-Preserving Synthetic Smart Meters Data.}
\newblock Available from \url{https://arxiv.org/abs/2012.04475}.
\newblock Accessed: 2021-10-27.

\bibitem[{Kaloskampis \emph{et~al.}(2021)Kaloskampis, Joshi, Cheung, Pugh, and
  Nolan}]{significance}
Kaloskampis I, Joshi C, Cheung C, Pugh D, Nolan L (2021).
\newblock \enquote{Synthetic Data in the Civil Service.}
\newblock \emph{Significance}, \textbf{17}, 18--23.

\bibitem[{Karr \emph{et~al.}(2006)Karr, Oganian, Reiter, and
  Woo}]{Karr_Oganian_Reiter_Woo_2006}
Karr A, Oganian A, Reiter J, Woo M (2006).
\newblock \enquote{New measures of data utility.}
\newblock Technical report available from:
  \url{http://citeseerx.ist.psu.edu/viewdoc/download?doi=10.1.1.576.5002&rep=rep1&type=pdf}.
\newblock Accessed: 2016-12-12.

\bibitem[{Nowok \emph{et~al.}(2014)Nowok, Raab, and
  Dibben}]{synthpop_1.0-0_CRAN}
Nowok B, Raab GM, Dibben C (2014).
\newblock \emph{\pkg{synthpop}: Generating Synthetic Versions of Sensitive
  Microdata for Statistical Disclosure Control}.
\newblock \proglang{R}~package version~1.0-0,
  \urlprefix\url{https://CRAN.R-project.org/package=synthpop}.

\bibitem[{Nowok \emph{et~al.}(2016)Nowok, Raab, and Dibben}]{synthpop_JSS}
Nowok B, Raab GM, Dibben C (2016).
\newblock \enquote{\pkg{synthpop}: Bespoke Creation of Synthetic Data in
  \proglang{R}.}
\newblock \emph{Journal of Statistical Software}, \textbf{74}(11), 1–26.
\newblock \doi{10.18637/jss.v074.i11}.
\newblock
  \urlprefix\url{https://www.jstatsoft.org/index.php/jss/article/view/v074i11}.

\bibitem[{Raab \emph{et~al.}(2017{\natexlab{a}})Raab, Nowok, and
  Dibben}]{Raab_jpc}
Raab G, Nowok B, Dibben C (2017{\natexlab{a}}).
\newblock \enquote{Practical Data Synthesis for Large Samples.}
\newblock \emph{Journal of Privacy and Confidentiality}, \textbf{7}, 67--97.
\newblock
  \urlprefix\url{https://journalprivacyconfidentiality.org/index.php/jpc/article/view/407}.

\bibitem[{Raab \emph{et~al.}(2017{\natexlab{b}})Raab, Nowok, and
  Dibben}]{Raab_Nowok_Dibben_2017}
Raab GM, Nowok B, Dibben C (2017{\natexlab{b}}).
\newblock \enquote{Guidelines for Producing Useful Synthetic Data.}
\newblock Available from \url{http://arxiv.org/abs/1712.04078}.

\bibitem[{Read and Cressie(1988)}]{Read_Cressie_1988}
Read T, Cressie RC (1988).
\newblock \emph{Goodness-of-Fit Statistics for Discrete Multivariate Data}.
\newblock Springer-Verlag.

\bibitem[{Snoke \emph{et~al.}(2018)Snoke, Raab, Nowok, Dibben, and
  Slavkovic}]{snoke_et_al_2018}
Snoke J, Raab G, Nowok B, Dibben C, Slavkovic A (2018).
\newblock \enquote{General and Specific Utility Measures for Synthetic Data.}
\newblock \emph{Journal of the Royal Statistical Society B}, \textbf{181}(3),
  663--668.

\bibitem[{Voas and Williamson(2001)}]{Voas_Williamson_2001}
Voas D, Williamson P (2001).
\newblock \enquote{Evaluating Goodness-of-Fit Measures for Synthetic
  Microdata.}
\newblock \emph{Geographical and Environmental Modelling}, \textbf{5},
  177--200.

\bibitem[{Woo \emph{et~al.}(2009)Woo, Reiter, Oganian, and Karr}]{wooProp}
Woo MJ, Reiter JP, Oganian A, Karr AF (2009).
\newblock \enquote{Global Measures of Data Utility for Microdata Masked for
  Disclosure Limitation.}
\newblock \emph{Journal of Privacy and Confidentiality}, \textbf{1}, 111--124.

\end{thebibliography}

\newpage

\begin{appendix}

\section{Details of utility measures} \label{app:detmeas}

Table~\ref{tab:detmeas} gives details of each of the utility measures mentioned in Section~\ref{sec:meas}, calculated from propensity score methods by \code{utility.gen()} and from tables by \code{utility.tab()}, using the notation given in Table~\ref{tab:notation}.   

\begin{table}[ht]
	\centering
	\begin{tabular}{ll}
		\hline
		Notation      &  Description\\ 			
		\hline
		$n_1, n_2, N$ &	Number of records in the original, synthesized, and combined data.\\
		$c$           &	Proportion of synthesized rows in the combined data $= n_2/N$.\\
		\multicolumn{2}{l} {\emph{Propensity score methods}}\\
		$t_i$         & Indicator variable, $i = 1, ..., N$, taking the value 1 for rows from\\
		              & synthetic data and 0 for rows from original data.\\
		$df$          & Degrees of freedom constrained by fitting models with a fixed number\\
		              & of parameters.\\
		$\hat{p}_i$   & Predicted probabilities, $i = 1, ..., N$, that row comes from synthetic data.\\\\
		\multicolumn{2}{l} {\emph{Tabular methods}}\\
		$o_1,...,o_j,...,o_k$ & Counts of original data in each cell.\\
		$s_1,...,s_j,...,s_k$ & Counts of synthetic data in each cell.\\
				$k$  & Total number of cells in tables to be compared, where $o_j + s_j >0$.\\
		$\hat{p}_j=s_j/(s_j+o_j)$ & Predicted probability that the $(s_j+o_j)$ counts in the $j^{th}$ cell are from\\
		                          & the synthetic data.\\
		\hline
		
	\end{tabular}
	\caption{\label{tab:notation}Notation used to describe utility measures.}
	
\end{table}

\begin{table}[h]
	\centering
	\begin{tabular}{lll}
		\hline
		Measure    &  Formula  &  Null expectation\\
		\hline
		\multicolumn{2}{l}{\textit{From propensity score methods}}\\
		$pMSE$ & $\sum_i(\hat{p_i}-c)^2 / N$                                                    & $df c(1-c)^2/N$ \\
		$PO50$ & $100 \sum_i{[t_i(\hat{p}_i > c)  + (1 -t_i)(\hat{p}_i < c) ]}/ \sum(\hat{p_i} \neq c) - 50$\\ 
		$SPECKS$   & $sup_e |F_{t=0}(\hat{p_i}) - F_{t=1}(\hat{p_i})|$ \\
		$U$    & Sum of ranks of $\hat{p_i}$, where $t_i = 1$ in the ordering of $\hat{p_i}$\\
		\multicolumn{2}{l}{\textit{From tabular methods}}\\
		$VW$   & $\sum_{j=1}^{k, (o_j + s_j) > 0}{{[s_j-o_jc/(1-c)]^2}/[c(o_j+s_j)]} $          & $k-1$\\
		$pMSE$ & $\sum_j[(o_j + s_j)(s_j/(o_j + s_j)-c)^2] / N$                                 & $(k-1)c(1-c)^2/N$	\\
		$G$    & $2\sum_{j=1}^{k,~o_j>0~ \&~ s_j>0} \{ (s_j/n_2) \log_e[(s_j/n_2)/(o_j/n_1)\}$  & $k-1$\\
 		$FT$   & $4\sum_{j=1}^{k}{[\sqrt{s_j} - \sqrt{o_jc/(1-c)}]}^2$                          & $k-1$\\ 
 		$dBhatt$  & $\sqrt{1 - \sum_{j=1}^{k}{\sqrt{(s_j/n_1)(o_j/n_2)}}}$ \\ 
		$JSD$  & $\sum_{j=1}^{k} \{(s_j/n_2) \log_2[2 (s_j/n_2)/((s_j/n_2) + (o_j/n_1))] $      &  $(k-1)log(2)/(2 N)$ \\
		            & $ + (o_j/n_1) \log_2[2 (o_j/n_1)/((s_j/n_2) + (o_j/n_1))\}/2$\\
		$PO50$ & $100 \sum_j{[t_j(\hat{p}_j >= 0.5)  + (1 -t_j)(\hat{p}_j < 0.5) ]}/N$\\
		$SPECKS$   &  $sup_e |F_{o_j}(\hat{p_j}) - F_{s_j}(\hat{p_j})|$, where $F_{w}(x)$ is the\\
		       & CDF of $x$ weighted by $w$.\\
		$MabsDD$  &  $\sum_{j=1}^{k} |o_j/n_1 - s_j/n_2|$    \\
		$WMabsDD$ &  $\sum_{j=1}^{k} | s_j - o_j n_2/n_1 | /\sqrt{2 c(o_j+s_j)/ \pi}$            & $k -1$\\
		$U$    & Sum of ranks of $\hat{p_j}$ weighted by $o_j$  in the\\
		       & ordering  of $\hat{p_j}$ weighted by $o_j + s_j$ \\
		\hline
	\end{tabular}
	\caption{\label{tab:detmeas} Utility measures from propensity score methods and from tabular methods.}
\end{table}
It is easy to show from the formulae\footnote{Calculating $VW$ from tables is equivalent to fitting a saturated logistic model with $df = k - 1$} in Table~\ref{tab:detmeas} that $pMSE = VW c(1-c)^2/N$. Note that the range of values and the Null expectation also correspond. When a table contains cells where both the original and synthetic counts are zero, the corresponding logistic model will have its $df$ reduced because some parameters will be aliased. 

Table~\ref{tab:detmeas} also gives range of possible values and the Null expectation of a model with a fixed number of parameters for each measure, when the latter is known. Note that the expressions for the Null expectations apply for large samples. For measures  from \code{utility.tab()} we might expect this to apply to each cell of the tables compared. For those derived from \code{utility.gen()} we might expect that it would be sufficient for the residual degrees of freedom $df$ from the fitted model to be large. This is investigated empirically in Appendix~\ref{app:evalmeas}. 

The utility measures can be scaled either by the maximum of their range or by their Null expectation. In practice most measures give values that are much closer to zero than to the maximum of their range, so scaling by the maximum 
gives very small ratios. Scaling by the Null expectation is more satisfactory. When no expression is available for the Null expectation then it will need to be computed by replication methods. Where utility measures are being used to compare altered data that are not created by the type of methods implemented in \pkg{synthpop}, for example by a method to provide differentially private micro-data, then scaling by the Null expectation can be considered as utility relative to the stochastic variation about the model used for the propensity score.

\section{Evaluation of utility measures}\label{app:evalmeas}
The most important aspect of a utility measure is its ability to identify differences between the synthetic and original data distributions. Another desirable feature is that the measure should have a known Null expectation that gives the results that would be obtained from a correct generative model. An empirical study, described below, assesses these two aspects for the measures given in Table~\ref{tab:detmeas}, but selecting only one from each set of measures that were shown in Section~\ref{sec:meas} to be linearly related.

New methods based on log-linear models were introduced into \pkg{synthpop} from Version~1.5-0. These models synthesize a group of variables together rather than proceeding via conditional distributions. There are two choices of model: either a log-linear model specified by its marginals (method \code{"ipf"}) or a saturated log-linear model corresponding to a complete cross-tabulation of all variables (method \code{"catall"}). They apply to categorical data, but an option allows continuous data to be grouped.  Because the computations involve creating large tables, the memory available limits the number of variables that can be included together. When used as part of a large synthesis these methods are usually used for just the first few important variables, and other variables then synthesised from conditional distributions. But if synthetic data are generated from a few categorical variables, all with method \code{"catall"}, the data will have been generated from  a  model that would be ``correct'' from the perspective of any of the models (a), (b) or (c) in Table~\ref{tab:models}.  In contrast, we can generate an ``incorrect'' synthesis from a model where there are no relationships between the variables by using the method \code{"sample"} for all variables taking independent bootstrap sample from all values. The difference in the utilities from an ``incorrect'' to a ``correct'' synthesis, scaled by the standard deviation of the measure from the ``correct'' synthesis is a measure of the power of the measure to detect differences in the distributions.

The code below was used to produce the results for 1,000 syntheses for subsets of variables from the data \code{ods} used in Section~\ref{sec:intro}. Numeric variables in \code{ods} were first grouped into cateories and then data sets consisting of the first $n$ variables, with $n$ ranging from two to six, were synthesized by two methods: the first using method \code{"catall"} for all variables (a ``correct'' model) and the second using method \code{"sample"} for all variables (a poor model, not attempting to fit any relationships between variables). Each variant produced 1,000 syntheses. The mean difference between each utility measure from the poor model and that from the ``correct'' model, was calculated and standardized by the standard deviation of the ``correct'' model. Results are in Table~\ref{tab:power} and discussed in Section~\ref{sec:meas}. As we saw above the three measures $VW$, $FT$ and $JSD$ have the best power. The measure $WMabsDD$ also has good power, especially for larger tables. The measures $SPECKS:MabsDD:PO50$ and $U$ do somewhat less well. The likelihood ratio $G$ loses power for larger sparse tables due to the relatively few cells that contribute\footnote{as we can see from $dfG$ in Table~\ref{tab:power}.}. This is an extreme example, ignoring all dependencies between the variables, so the utility measures are all very high. The same code as given above can be used to compare the power of syntheses by a CART model with the correct model. For two variables the power of the utility measures were all zero, because in every case the CART model was able to reproduce the two-way table. For larger tabulations the empirical power was lower, but the relative patterns were similar to those in Table~\ref{tab:power}. For example, for 4 variables the empirical power was 16 for $pMSE:VW$ compared to 12 for $SPECKS:MabsDD:PO50$.

\begin{CodeChunk}
	\begin{CodeInput}
R> m <- 1000
R> ods_cat <- numtocat.syn(ods, cont.na = list( income = -8), style.groups = "quantile", 
catgroups = 5)$data
R> syn_bad   <- syn(ods_cat, method = "sample",  print.flag = FALSE, m=m, seed = 12345)

R> calc_power <- function(nvars,m = m) {
	syn_good <- syn(ods_cat[,1:nvars] ,  method = "catall", print.flag = FALSE, 
	m = m, seed = 12345)
	u_good <- utility.tab(syn_good, ods_cat,  vars = names(syn_good$syn[[1]]),
	print.flag = FALSE) 
	u_bad  <- utility.tab(syn_bad  , ods_cat, vars = names(syn_good$syn[[1]]),
	print.flag = FALSE) 
	results_bad <- data.frame( pMSE = u_bad$pMSE,  FT =u_bad$FT,  JSD =u_bad$JSD, 
	SPECKS = u_bad$SPECKS,  U = u_bad$U,  WMabsDD = u_bad$WMabsDD,  G = u_bad$G, 
	df = u_bad$df,  dfG = u_bad$dfG)
	results_good <- data.frame( pMSE = u_good$pMSE,  FT =u_good$FT,  JSD =u_good$JSD, 
	SPECKS = u_good$SPECKS, U = u_good$U, WMabsDD = u_good$WMabsDD, G = u_good$G, 
	df = u_good$df,  dfG = u_good$dfG )
	power <- (apply(results_bad[,1:7],2,"mean") - apply(results_good[,1:7],2, "mean"))
	/sqrt(apply(results_good[,1:7],2,"var"))
	list(power = c(power,df_good = median(results_good[,8]),dfG_good = 
	median(results_good[,9]), df_bad = median(results_bad[,8]), dfG_bad = 
	median(results_bad[,9])),
	expect = apply(data.frame(S_pMSE = u_good$S_pMSE,S_FT = u_good$S_FT,
	S_JSD = u_good$S_JSD, S_WMabsDD = u_good$S_WMabsDD,S_G = u_good$S_G), 2 ,"mean"))
}
R> pe <- as.list(2:6)
R> for (i in 2:6) {pe[[i-1]]<- calc_power(i,m)}
R> table1 <- rbind(pe[[1]]$power,pe[[2]]$power,pe[[3]]$power,
pe[[4]]$power,pe[[5]]$power)
R> table5 <- rbind(pe[[1]]$expect,pe[[2]]$expect,pe[[3]]$expect,
pe[[4]]$expect,pe[[5]]$expect)

	\end{CodeInput}
\end{CodeChunk}

\section{Methods for NULL distribution of utility measures}\label{app:evalNULL}

The code in the section above used to produce Table~\ref{tab:power} also computes utility measures standardized by their Null expectation, for those measures where it can be calculated from their Null expectation. If the Null expectation is valid these should give average values of 1.00. The results in Table~\ref{tab:expect} show that most are close enough to 1.00 for practical use but for sparse tables $S\_G$ is underestimated and $S\_JSD$ overestimated.

\begin{table}[h]
	\centering
	\begin{tabular}{rrrrrrr}
		\hline
		Number       & Mean & Mean & Mean & Mean & Mean \\
		of variables	& $S\_pMSE$ & $S\_FT$ & $S\_JSD$ &  $S\_WMabsDD$ & $S\_G$ \\
		\hline
		2 & 1.01 & 1.01 & 1.05 & 1.04 & 1.01 \\ 
		3 & 1.01 & 1.02 & 1.06 & 1.01 & 1.01 \\ 
		4 & 1.03 & 1.15 & 1.12 & 1.00 & 0.96 \\ 
		5 & 1.06 & 1.51 & 1.29 & 1.00 & 0.82 \\ 
		6 & 1.06 & 1.60 & 1.33 & 1.00 & 0.74 \\
		\hline
	\end{tabular}
	\caption{\label{tab:expect} Average standardised utility measures for a ``correct'' categorical syntheses calculated by \code{utility.tab()} with the code given above using 1,000 syntheses of each data set.} 
\end{table}

These results only apply to some measures and for complete syntheses. More importantly, they require models, such as logit models, with a fixed number of parameters. For adaptive models such as CART, the expectation must be computed by resampling methods. \cite{snoke_et_al_2018} describe the two possible resampling methods: permuting the propensity score vector while leaving the indicator variable fixed (perm) and creating multiple synthetic data sets and computing the utility from pairs of syntheses (pairs). Either method can be used for the $pMSE$ utility, but for $SPECKS$, $PO50$ and $U$ only the pairs method can be used. The permutation method gives results that are completely wrong for these measures. This can be understood by realizing that the distribution of these statistics depend on  the propensity score calculations. Just permuting the indices does not produce CDFs where fitting the propensity score model forces the CDF for the synthetic data to lie above that for the original. Table~\ref{tab:last} gives the results from the code below to evaluate the resampling methods by the pairs method.

\begin{table}[h]
	\centering
	\begin{tabular}{cccccc}
		\hline
		& \multicolumn{3}{c}{ $S\_pMSE:S\_VW$} & $S\_SPECKS:S\_MabsDD:PO50$ & $S\_U$ \\
		variables	& calculated$^a$   & permutation$^b$ & pairs$^c$ & pairs$^c$  & pairs$^c$ \\
		\hline
2 & 1.01 & 0.88 & 0.95 & 1.33 & 1.98 \\ 
3 & 1.01 & 0.83 & 0.79 & 1.24 & 1.95 \\ 
4 & 1.03 & 0.72 & 0.71 & 1.18 & 1.93 \\ 
5 & 1.06 & 0.66 & 0.66 & 1.16 & 1.91 \\ 
6 & 1.06 & 0.64 & 0.66 & 1.15 & 1.90 \\
		\hline
		
		\multicolumn{6}{l}{$^a$ from the 1000 correct syntheses }\\
		\multicolumn{6}{l}{$^b$ from 50 permutations of each of 1000 syntheses}\\
		\multicolumn{6}{l}{$^c$ from the 1035 pairs from 46 syntheses}\\
		\hline
	\end{tabular}
	\caption{\label{tab:last} Mean standardized utility measures for correct syntheses by different methods.}
\end{table}\textbf{}

\begin{CodeChunk}
	\begin{CodeInput}
R> calc_power2 <- function(nvars) {
	syn_good1 <- syn(ods_cat[,1:nvars] ,  method = "catall", print.flag = FALSE, 
	m = 1000, seed = 12345)
	u_good1 <- utility.gen(syn_good1, ods_cat[,1:nvars],  
	vars = names(syn_good1$syn[[1]]), resamp.method = "perm", 
	utility.stats = c("S_pMSE"), print.flag = FALSE) 
	syn_good2 <- syn(ods_cat[,1:nvars] ,  method = "catall", print.flag = FALSE, 
	m = 46, seed = 12345)
	u_good2 <- utility.gen(syn_good2, ods_cat, 
	vars = names(syn_good2$syn[[1]]),resamp.method = "pairs",
	utility.stats = c("S_pMSE","S_SPECKS","S_U"),  print.flag = FALSE)
	table6 <- c(S_pMSE1 = table5[nvars-1, 1], S_pMSE2 = mean(u_good1$S_pMSE), 
	S_pMSE3 = mean(u_good2$S_pMSE), S_SPECKS = mean(u_good2$S_SPECKS),
	S_U = mean(u_good2$S_U))
}
R> pe <- as.list(2:6)
R> for (i in 2:6) {pe[[i-1]]<- calc_power2(i)}
R> table6 <- rbind(pe[[1]],pe[[2]], pe[[3]], pe[[4]],pe[[5]])
	\end{CodeInput}
\end{CodeChunk}

Results from replication methods are clearly less satisfactory than for measures that have a known Null distribution. However, given that we are mainly interested in diagnosing aspects of the synthesis with standardised measures above 10, they may not be entirely useless. These results, taken together, favour our recommendation to use $pMSE$ as the best utility measure, but with $WMabsDD$ having an equivalent performance. This second measure can be considered as the $L1-norm$ equivalent of $pMSE$ as an $L2-norm$, and so may have some advantage in terms of robustness to out liers.

\section[Equivalence of SPECKS, PO50 and MabsDD]{Equivalence of $SPECKS$, $PO50$ and $MabsDD$} 
\label{app:plot}

\begin{figure}[h]
	\centering
	\includegraphics[height=8cm,width=8cm]{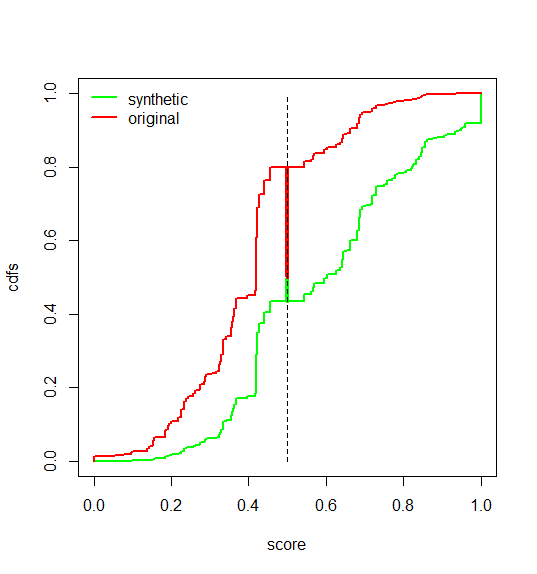}
	\caption{\label{fig:cumplot} Cumulative distributions of propensity scores for the uility evaluation by a CART   model with \code{utility.gen()} of a poor synthesis of \code{ods} data.}
\end{figure}

This graphical proof of equivalence of $SPECKS$, $PO50$ and $MabsDD$ is for the case when the original and synthesised data sets have the same number of records ($N/2$). Figure~\ref{fig:cumplot} illustrates the cumulative distribution functions (CDFs) of the propensity scores for the original and synthetic data for a poor synthesis that does not model the correlations between the variables. The Kolmogorov-Smirnov statistic is the maximum distance between these CDFs, illustrated here. When both of the CDFs are evaluated at the same set of points, as is the case here, the maximum distance between the CDFs is at a score value of 0.5. The CDF for the original data at a score of 0.5 is the proportion of the original data correctly predicted as not being synthetic, and hence the solid red vertical line is the proportion correct  above 0.5. Similarly the green vertical line is the proportion over 0.5 correctly predicted for the synthetic data. The distances on the graph have denominators of $N/2$ while $PO50$ has a denominator of $N$, so see that $SPECKS = 2 PO50/100$. When the scores are calculated from tables, the points contributing to the CDFs with scores below 0.5 will refer to table cells where the counts of original records exceeds the synthetic counts. The opposite is true for scores above 0.5. Thus the $MabsDD$ is exactly twice $SPECKS$. These results hold for  the original and synthesised data sets that have the same number of records. We have not yet explored the case of unequal numbers. Also these results may only be approximate when the score values at which the synthetic and original data are evaluated do not coincide.  

\end{appendix}


\end{document}